\documentclass[12pt,letterpaper]{article}
\usepackage{local}
\usepackage{graphicx}
\usepackage[T1]{fontenc}
\usepackage{amsmath}
\usepackage{amsfonts}

\begin{document}

\title{Controlling light polarization by swirling surface plasmons} 

\author{Mengjia Wang,$^{1}$ Roland Salut,$^{1}$ Huihui Lu,$^{2\ast\ast}$ Miguel-Angel Suarez,$^{1}$\\ Nicolas Martin,$^{1}$ and Thierry Grosjean,$^{1\ast}$}

\affiliation{$^{1}$FEMTO-ST Institute UMR 6174, University of Bourgogne Franche-Comte - CNRS - Besancon, France\\
$^{2}$ Guangdong Provincial Key Laboratory of Optical Fiber Sensing and Communications, Department of Optoelectronic Engineering, Jinan University, Guangzhou 510632, China}

\address{Email: *thierry.grosjean@univ-fcomte.fr and **thuihuilu@jnu.edu.cn}

\begin{abstract}
\textbf{Light polarization is a key aspect of modern optics. Current methods for polarization control utilize birefringence and dichroism of anisotropic materials or of arrays of anisotropically shaped nanostructures. Based on collective optical effects, the resulting components remain much larger than the wavelength of light, which limits design strategies. Here, we present a travelling-wave plasmonic antenna that overcomes this limit using a gold-coated helical nanowire non-radiatively fed with a dipolar aperture nanoantenna. Our non-resonant hybrid nanoantenna enables tightly confined circularly polarized light by swirling surface plasmons on the subwavelength scale and taking advantage of optical spin-orbit interaction. Four closely packed circularly polarized light sources of opposite handedness and tunable intensities are demonstrated. By reaching near-field interaction between neighboring nanoantennas, we obtain a highly miniaturized wave plate whose polarization properties have never previously been demonstrated.} 
\end{abstract}

\maketitle

\newpage

\fontsize{12}{20} \selectfont

A wide variety of optical applications and techniques require control of light polarization. Traditionally, manipulation of the polarization vector of light is realized with bulky optical elements, which utilize birefringent, dichroic or optically active media. This field has recently experienced extraordinary advances with the emergence of plasmonics, which has provided new mechanisms for light--matter interactions. Surface plasmon (SP) resonances in subwavelength metallic structures have laid the groundwork for metamaterial research, leading to ultrathin circular polarizers \cite{gansel:sci09,zhao:natcom12,kaschke:aom15} and waveplates \cite{yu:nl12,zhao:prb11,zhao:nl13,zhu:ol15,ding:acsnano15,drezet:prl08} by locally tailoring the phase of light \cite{yu:sci11,drezet:prl08} or generating chirality, optical activity or circular dichroism \cite{plum:prb09,hendry:natnano10,schaferling:prx12,cui:nl14,decker:ol10,ren:natcom12,kuwata:prl05,rogacheva:prl06,garoli:nl16}. However, metamaterials are restricted to areas larger than the wavelength of light, as they rely on collective optical effects on arrays of resonant nanostructures. Therefore, manipulation of light polarization with components and wave plates of negligible footprint, i.e., of subwavelength extent, remains a significant challenge \cite{khoo:ol11}.

A characteristic of travelling SPs guided along curved trajectories is their ability to acquire extrinsic orbital angular momentum (EOAM) \cite{bliokh:natphot15,lefier:nl18} and to induce, by leakage, free-space light emission \cite{balanis:book}. For sharp curvatures, the EOAM of the SP mode can match the spin angular momentum (SAM) of free-space propagating photons, thereby opening a route towards localized light emissions of controlled helicity, owing to optical spin-orbit interaction (SOI) \cite{bliokh:natphot15}. On the basis of this EOAM-to-SAM transfer, we have generated a plasmonic helical antenna (PHA) to produce circularly polarized directional light on the subwavelength scale through a swirling plasmonic effect. Our optical antenna differs from existing helical plasmonic structures \cite{gansel:sci09,schaferling:acsphot14,wozniak:ox18,passaseo:aom17} by its non-resonant nature, thus extending the concept of travelling-wave helical antenna to optics \cite{balanis:book,kraus:book}. With four closely packed PHAs of opposite handedness, we demonstrate deterministic manipulation of light polarization with a device of subwavelength lateral extent, while achieving properties in polarization transformation so far unattainable. 


Our plasmonic antenna consists of a narrow gold wire wound up in a screw-like shape forming a tiny helix (Fig. 1(a)). The gold-coated wire sustains a cutoff-free axially symmetric travelling SP, known as the $TM_0$ mode \cite{novotny:pre94} (Fig. 1(b)). It is locally fed with the dipolar mode of a rectangular aperture nanoantenna that perforates a 100 nm thick gold layer right at the helix's pedestal.  An incident wave on the back of the aperture is transmitted as a subdiffraction guided SP, which is non-radiatively converted into the wire mode of the helix. The contact between the aperture and the helix's pedestal ensures efficient near-field coupling between the two plasmonic structures. To identify the travelling-wave nature of the antenna, we showed the intensity of the current along the metallic wire of a four-turn PHA (Fig. 1(d)). The thus-depicted mode closely resembles a travelling wave, as no clearly marked current nodes are evidenced.  

\begin{figure}[ht!]
\centering
\includegraphics[width=0.65\linewidth]{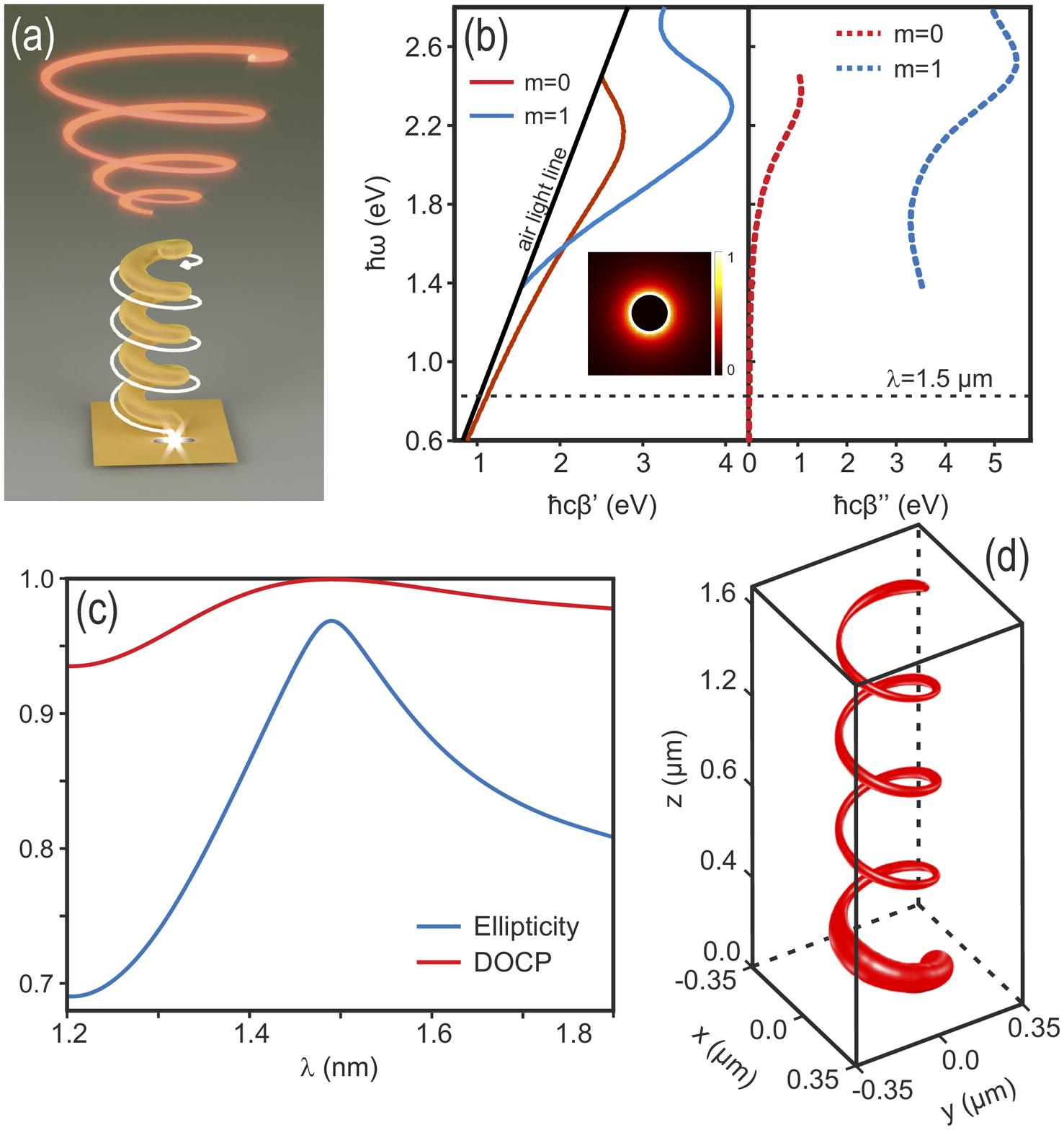}
\caption{\textbf{Helical plasmonic antenna as a circularly polarized subwavelength source.} \textbf{a}, Schematics of the PHA and its operation principle. End-fire excitation is represented with a white spike. Under a curved trajectory along the helix, SPs acquire EOAM and are simultaneously released as freely propagating waves (white arrow). Part of the mode leakage re-excites the plasmon wire mode, thus participating in the swirling plasmonic effect. \textbf{b}, Dispersion relations of the $m=0$ and $m=1$ modes of a gold-coated carbon wire (105-nm diameter carbon wire, 25-nm thick gold coating). Energy is plotted versus $\hbar c \beta'$ and $\hbar c \beta"$. At $\lambda$=1.5 $\mu$m, only the $m=0$ mode is guided. Figure inset: intensity plot of the $m=0$ mode for $\lambda$=1.5 $\mu$m. \textbf{c}, Spectra of the ellipticity factor and DOCP of the PHA emission. \textbf{d}, Amplitude of the electric current distribution along the gold-coated carbon wire, at $\lambda$=1.5 $\mu$m.} 
\end{figure}

In the course of propagation, the plasmon wire mode acquires EOAM oriented along the helix axis (0z). At the same time, the SPs are released as free-space propagating waves carrying an SAM of $\pm 1$ per photon (in $\hbar$ units). Part of the emitted waves interacts with the helix and re-excites the plasmon wire mode, thereby participating in the swirling plasmonic effect. This travelling wave property arises from the fact that our PHA is a chiral plasmonic waveguide operating near cutoff. The degree of circular polarization (DOCP) of the emitted waves refers to the distribution of photons prepared in the spin states $+1$ and $-1$. The DOCP is defined as $|I_{RCP}-I_{LCP}|/(I_{RCP}+I_{LCP})$ where $I_{RCP}$ and $I_{LCP}$ stand for the intensities of the right and left circularly polarized components of the antenna emission, respectively\cite{yu:nl12}. A PHA designed for operation at $\lambda$=1.5 $\mu$m has been predicted to emit light with polarization ellipticity and a DOCP peaking at 0.97 and 0.999, respectively (Fig. 1(c)). 


Our fabrication of the corresponding structures started with the growth of carbon helices by focused ion beam-induced deposition (FIBID) \cite{esposito:nl16} on a 100 nm thick gold film covering a glass substrate. The carbon helices were then coated with a thin layer of gold. The PHA was terminated by focused ion beam (FIB) milling of a single rectangular aperture nanoantenna in contact with the helix pedestal and outside the winding area of the plasmonic wire (see Supplementary Fig. 5). Fig. 2(a) and the inset of Fig. 2(b) display scanning electron microscopy (SEM) images of a resulting structure. The PHAs were back-illuminated with polarized light from a tunable laser at telecommunication wavelengths, the antenna emissions were measured and their polarization state was analyzed (see Methods). The observed polarization properties (Fig. 2(b) and 2(c)) agree well with the theoretical model. As anticipated in Fig. 2(b), a gold coating thinner than predicted may explain the noticeable redshift in the experimental spectra with respect to the numerical simulation. 

\begin{figure}[ht!]
\centering
\includegraphics[width=0.85\linewidth]{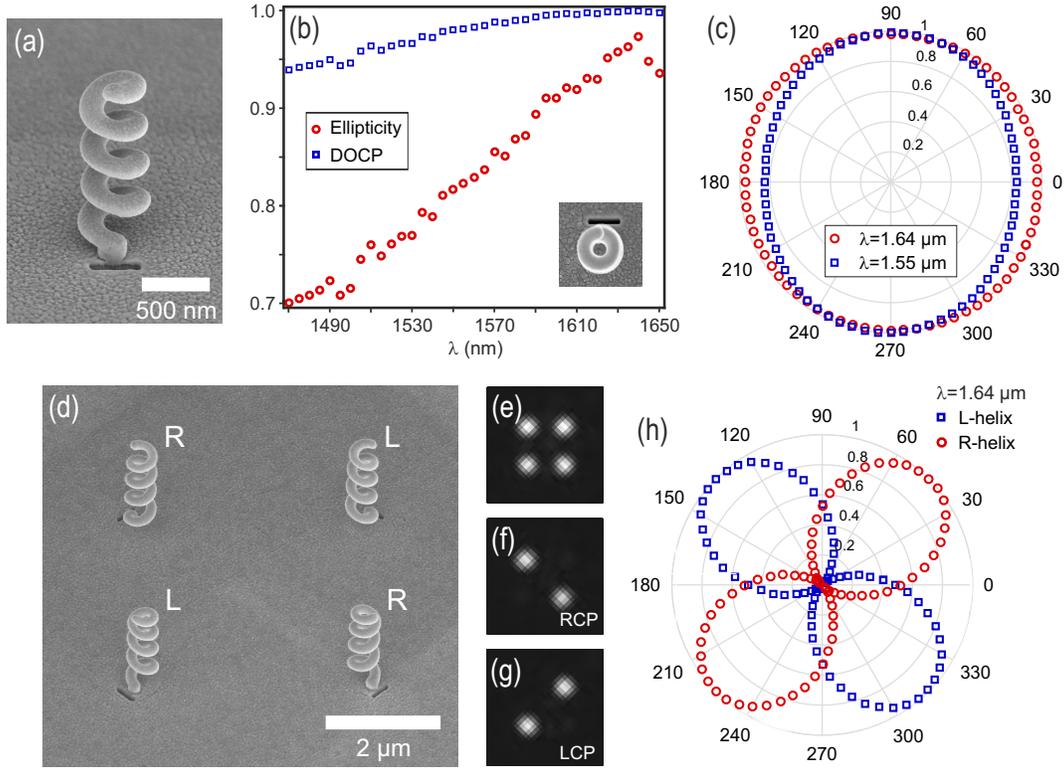}
\caption{\textbf{Circularly polarized optical emission from fabricated single plasmonic helical antennas.} \textbf{a}, Scanning electron micrograph of a PHA. \textbf{b}, Experimental spectra of the ellipticity factor and DOCP of the PHA emission. Inset: top view of \textbf{a}. Image size: 1 $\mu$m. Solid curves: theoretical spectra (ellipticity factor and DOCP) with a 10 nm thick gold layer covering the carbon helix. \textbf{c}, State-of-polarization analysis at $\lambda$=1.55 $\mu$m and $\lambda$=1.64 $\mu$m. \textbf{d}, Scanning electron micrograph of two couples of PHAs of opposite handedness and orthogonal aperture nanoantennas. The right- and left handed PHAs are identified with the letters R and L, respectively. \textbf{e-g}, Far-field optical images of the four-PHA device in \textbf{d}, with \textbf{f}, a right-handed and \textbf{g}, a left-handed circular analyser in front of the camera. \textbf{h}, Helicity analysis of two PHAs of opposite handedness. The measurement is conducted by placing a rotating quarter-wave plate followed by a fixed polarizer in front of a detector and measuring the transmitted power.}
\end{figure}

When such HPAs of opposite handedness are arranged with a spacing of a few microns, the overall structure transmits light in the form of a small array of background-free right- and left-handed circularly polarized sources (Figs. 2(d)-2(h)). Moreover, by using PHAs with various orientations of the feed apertures, the relative intensities of these point-like emissions become controllable by the polarization of the incident light. We can therefore create ultracompact optical architectures made of tiny circularly polarized directional light sources (Supplementary Fig. 6) with arbitrary handedness and tunable intensities, thereby obtaining unprecedented integrated devices for manipulating light polarization on a small scale. The enhancement in the ellipticity factor with the number of turns of the helix reveals the swirling plasmonic effect as the source of circular polarization (Supplementary Section 1). The chiral nature of the PHA generates circular dichroism (Fig. 3). However, circular dichroism occurs only when the antenna is illuminated from the top (i.e., in collection mode; see Fig. 3(a)). The measured indicates a transmission process mainly governed by near-field coupling between the helix and the rectangular aperture nanoantenna. Such near-field coupling is confirmed in Supplementary Section 2. The PHAs are thus also appealing tools to discriminate right and left circular polarizations, for instance, to locally measure the DOCP of light and perform polarimetry on the subwavelength scale \cite{martinez:sci18}. 

\begin{figure}[ht!]
\centering
\includegraphics[width=0.8\linewidth]{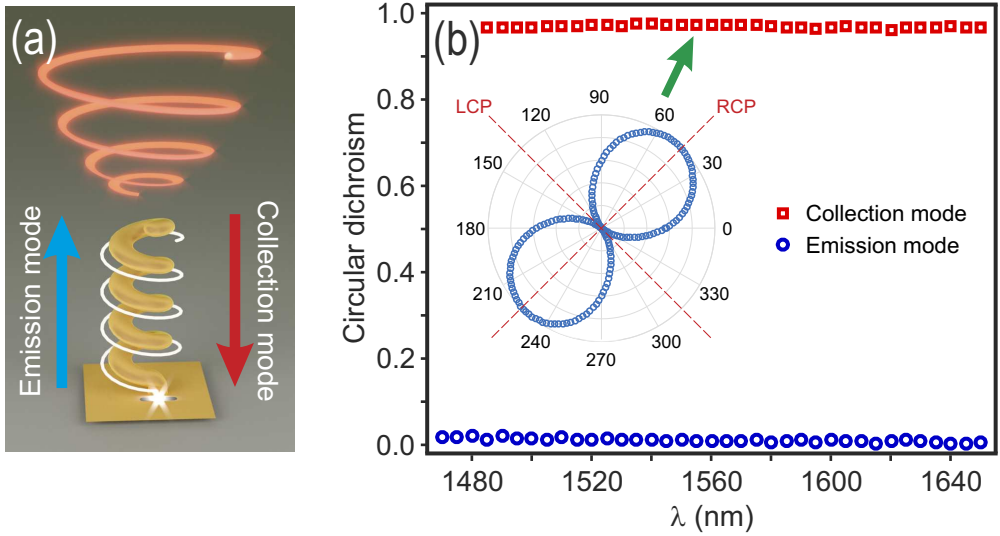}
\caption{\textbf{Plasmonic helical antenna and circular dichroism.} \textbf{a}, Schematics of the two operation modes of the PHA, involving light wave propagation in two opposite directions along the antenna axis. \textbf{b}, Circular dichroism spectra measured with a left-handed PHA operating in emission and collection modes. Circular dichroism is defined as $(I_{RCP}-I_{LCP})/(I_{RCP}+I_{LCP})$, where $I_{RCP}$ and $I_{LCP}$ stand for the emission intensities of the PHA, with illumination by right and left circularly polarized light, respectively. Emission intensity is measured from the helix, in air (in emission mode), or from the rectangular aperture nanoantenna through the substrate (in collection mode). Figure inset: helicity analysis in emission mode at $\lambda$=1.55 $\mu$m.}  
\end{figure}


A more complex polarization response can be achieved with a spacing between the PHAs that is smaller than the wavelength, resulting in the coupling of the light emission processes created by individual plasmonic antennas. We consider two couples of right and left PHAs with antennas of opposite handedness that are spaced at a distance of 560 nm apart, and are made up of orthogonal apertures (Figs. 4(a) and 4(b)). With this geometry, far-field emission shows a single spot regardless of the incident polarization (inset of Fig. 4(b)). Figs. 4(c) and 4(e) compare the measured and calculated tilt angles $2\psi$ and ellipticity angles $2\chi$ of the outcoming polarization (Poincare sphere approach) as a function of the direction angle $\theta$ of the incident linear polarization, at two different wavelengths (1.61 $\mu$m and 1.47 $\mu$m), respectively. The measured curves in Figs. 4(c) and 4(e) reveal the theoretically anticipated tuning of the angular spacing $\Delta \theta$ between the two right- and left-handed outcoming circular polarizations. Whereas $\Delta\theta$ is fixed at 90$^{\circ}$ with conventional phase retardation plates, it is here equal to 69$^{\circ}$ at $\lambda$= 1.61 $\mu$m and decreases down to 51$^{\circ}$ at $\lambda$=1.47 $\mu$m. Such tunability in polarization manipulation is not standard at all. It arises from the ability of our structure to generate circular polarizations from the combination of two elliptically polarized waves of parallel ellipticities, as shown experimentally in Figs. 4(d) and 4(f), and tunable intensities (see Supplementary Section 3). By spectrally detuning the PHAs, the outcoming polarization ellipticities are modified, thereby resulting in a controlled angular spacing $\Delta\theta$. 

\begin{figure}[ht!]
\centering
\includegraphics[width=0.9\linewidth]{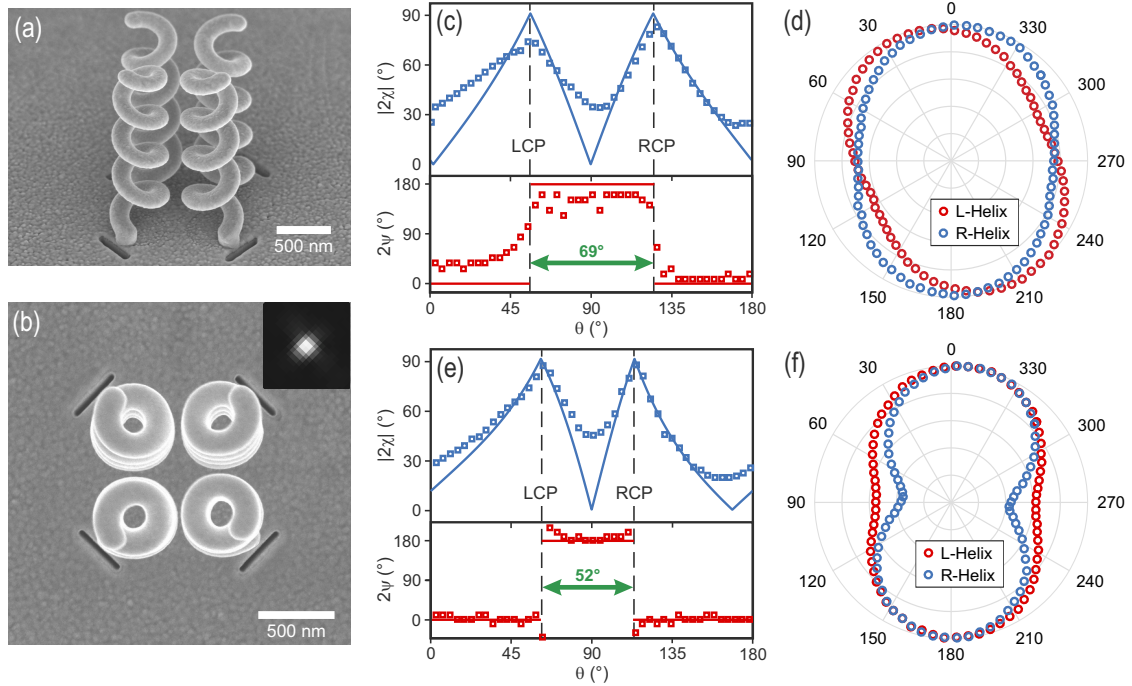}
\caption{\textbf{Subwavelength scale manipulation of light polarization by four coupled PHAs.} \textbf{a,b}, Scanning electron micrographs of the device. \textbf{a}, Angled view, \textbf{b}, top view.\textbf{c},\textbf{e}, Polarization angles $2\chi$ and $2\psi$  of the emitted field versus the polarization direction of incident linearly polarized light (Poincare sphere approach of polarization). The optical waves are impinging from the substrate at normal incidence with \textbf{c}, $\lambda$=1.61 $\mu$m and \textbf{e}, $\lambda$=1.47 $\mu$m. Owing to the oscillatory nature of light fields, the full set of incident linear polarizations is covered with a polarization angle $\theta$ ranging from 0 to 180 $^{\circ}$. Theoretical predictions are shown as solid lines (see Supplementary Section 3). \textbf{d},\textbf{f}, Polarization diagram of the antenna emission for incident polarization corresponding to $\theta$ equal to 45$^{\circ}$ and 135$^{\circ}$, leading to the selective excitation of the two couples of PHAs of opposite handedness: \textbf{d}, $\lambda$=1.61 $\mu$m, \textbf{f}, $\lambda$=1.47 $\mu$m. Near-field coupling between the PHAs of opposite handedness ensures parallel outcoming polarization ellipses for orthogonal incident linear polarizations.} 
\end{figure}

Based on the SOI of light, our method is versatile and leads to ultracompact plasmonic travelling-wave antennas. Swirling surface plasmons thus provide functionalities that have never previously been demonstrated. Taken as individual or coupled structures, the PHAs may pave the way towards highly integrated polarization-encoded optics, particularly for the generation and control of spin-encoded photon qubits in quantum information and optical spintronics. They may also enable new polarization-based optical functionalities for sensing or communications, which could include the unique electromagnetic and mechanical properties of 3D metal-coated helical wires.\\


\textbf{Simulations}

All the numerical simulations of the antenna emission process are realized using the 3D FDTD method. The plasmonic helix geometry considered in this study consists of a 105 nm diameter carbon wire wound up in the form of a four-turn corkscrew-type structure and covered with a 25 nm thick gold layer. The resulting helix has a 505 nm outer diameter and is 1.66 $\mu$m high.  It is positioned on a pedestal considered a 105 nm diameter and 100 nm high carbon rod whose cylindrical lateral side is covered with a 25 nm thick gold layer. The helix pitch angle is approximately 20.7$^{\circ}$. The helix pedestal lies on a 100 nm thick gold layer deposited onto a glass substrate. The rectangular aperture nanoantenna, with a width and length equal to 40 nm and 370 nm, respectively, is engraved in the metal layer. Its centre is located at x = y = 0. z = 0 corresponds to the upper surface of the gold layer that covers the glass substrate. To excite the PHA, a Gaussian beam (beam waist equal to 1.5 $\mu$m) impinges onto the rectangular aperture nanoantenna at normal incidence from the backside. 

The spectral response of the PHA is obtained with a Gaussian excitation described by a single temporal pulse. The time-varying electric field is calculated at a single cell located on the helix axis, 4 $\mu$m away from the end of the helix, along (0z). The spectra of the vector components $E_x$ and $E_y$ are calculated by Fourier transforming this result. From these results, the ellipticity factor is deduced as a function of the wavelength. The model used for the spectrum calculations consists of a volume spanning  4.55 $\mu$m in the x and y directions perpendicular to the longitudinal helix axis.  It extends 2 $\mu$m below the gold layer in the glass substrate and terminates 4.3 $\mu$m beyond the top of the helix in air.  All six boundaries of the computation volume are terminated with perfectly matched layers to avoid spurious unphysical reflections around the structure. The non-uniform grid resolution varies from 30 nm for portions at the periphery of the simulation to 5 nm within and near the helix and the aperture nanoantenna.

In all the calculations conducted in the continuous wave regime, the wavelength is 1.5 $\mu$m. The distribution of the current amplitude within the helix is plotted by integrating the simulated optical current density across the gold coating of the helix-shaped carbon wire for each curvilinear coordinate along the wire. The PHA geometry and mesh grid parameters remain unchanged. The computation volume spans 2.1 $\mu$m in the x and y directions. It extends 0.75 $\mu$m below the gold layer in the glass substrate and terminates 2.61 $\mu$m  beyond the top of the helix in air.\\ 

\noindent
\textbf{Fabrication}

PHAs have been fabricated in three steps using FIBID technology and FIB milling (Dual Beam SEM/FIB FEI Helios 600i). A helical carbon skeleton was fabricated by FIBID onto a commercial 100 nm thick gold film. For operation at 1.63 $\mu$m (experimentally), the geometrical parameters are a pitch angle of approximately 19.8$^{\circ}$, a radius of 155 nm and a pitch length of 350 nm. The structure was covered with a thin smooth layer of gold sputter-deposited by glancing angle deposition. Then, a 370 nm long and 40 nm large rectangular aperture nanoantenna was milled in contact with the helix pedestal (see Supplementary Fig. 5).\\

\noindent
\textbf{Characterization}

A schematic diagram of the experimental setup is represented in Supplementary Fig. 7. It is mounted onto a Nikon TE2000 inverted microscope. Light of tunable wavelength, ranging from 1.47 $\mu$m to 1.65 $\mu$m, emerges from a tunable laser source (Yanista Tunics-T100S) and is coupled to a single-mode polarization-maintaining fibre (P3-1500PM-FC-2, Thorlabs). It is collimated by an achromatic reflective fibre collimator (RC08APC-P01, Thorlabs) and focused onto the plasmonic structures with either a (25X, 0.4) microscope objective for the study of the single PHAs or a (4X, 0.1) objective for the four-coupled PHA structures. 

Two operation modes of the PHA are here investigated. They involve optical propagation in two opposite directions within the PHAs (Supplementary Fig. 7). In the emission mode , we consider the illumination of the rectangular aperture nanoantenna from the backside and the emission of circularly polarized light by the helix, whereas the collection mode corresponds to illumination of the helix from the front and emission of linearly polarized light from the rectangular aperture nanoantenna on the backside. Measurement in the collection mode are performed by inverting the illumination and collections benches in the optical characterization set-up.

Two parameters of the antennas are here investigated: their polarization diagram in emission mode and their circular dichroism in both emission mode and collection mode (Supplementary Fig. 7). In the former case, the polarization of the incident collimated wave is manipulated using a fixed polarizer (LPNIR100-MP2) and a half-wave plate (AHWP05M-1600, Thorlabs) positioned in between the collimator and the objective. The half-wave plate is mounted onto a motorized stage (PRM1Z8, Thorlabs) to be accurately rotated with respect to the polarizer. The plasmonic structures are imaged with an (50X, 0.65) infrared objective from Olympus (LCPlanN) coupled to an infrared camera (GoldEye model G-033, Allied Vision Technologies GmbH) and a proper field lens. To analyse the polarization state of the light emitted by our plasmonic antennas, either a rotating linear polarizer or a fixed polarizer coupled to a rotating quarter-wave plate (see inset) are inserted in front of the camera (Linear polarizer: LPNIR100-MP from Thorlabs, quarter-wave plate: AQWP05 M-1600 from Thorlabs, motorized stage: UE16CC from Newport). In Fig. 2, the spectra are obtained by analysing the state-of-polarization at each wavelength. The incident polarization is oriented perpendicularly to the long axis of the rectangular aperture nanoantenna, to excite its fundamental plasmon mode. Right- and left-handed circular analysers are obtained by orienting the fast-axis of the quarter wave plate at +45$^{\circ}$ and -45$^{\circ}$ relative to the polarizer axis, respectively. For circular dichroism measurements (Fig. 3), the combination of a linear polarizer and half-wave plate used in the illumination bench is replaced by a polarizer coupled to a quarter-wave plate (see insets of Supplementary Fig. 7). The detection bench consists of an objective, a field lens and a camera.\\


The authors are indebted to Sarah Benchabane for the acquisition of the FIBID module. This work is supported by the Labex ACTION program (contract ANR-11-LABX-01-01), the EIPHI Graduate School (contract ANR-17-EURE-0002), and by the French RENATECH network and its FEMTO-ST technological facility.


\newpage

\begin{center}
\huge{Supplementary Information}
\end{center}

\vspace{1cm}

\section{Helicity and swirling surface plasmons}

Supplementary Fig. \ref{fig:turns} represents the spectrum of the emission ellipticity factor of four different helices showing increasing number of turns. From one to four turns, the measured maximum ellipticity factor varies from 0.64 to 0.96, while undergoing spectral redshift. The enhancement of the ellipticity factor with the number of helix turns evidences the swirling plasmonic effect in the definition of optical helicity.

\section{End-fire excitation of the plasmonic helix}

Supplementary Fig. \ref{fig:end-fire} shows that polarization properties of the PHA rely on the end-fire excitation of its plasmonic helical wire. We studied the emission ellipticity factor of three antenna configurations involving three different excitation schemes from a rectangular aperture nano-antenna. The ellipticity factor peaking at a value larger than 0.92 when the aperture nano-antenna is in contact to the helix pedestal drops down to 0.73 when it is removed 185 nm away from the helix. When the rectangular nano-aperture antenna is rotated by 90$^{\circ}$, i.e., when the orientation of its dipolar plasmon mode becomes orthogonal to the radially-polarized wire mode of the plasmonic helix, the ellipticity factor of the PHA emission is reduced to 0.32. The 3D vectorial near-fields produced by the rectangular aperture nanoantenna may explain, in the latter case, the non-null coupling between the orthogonal nano-aperture and the helix. These results also show that the subdiffraction plasmonic wire mode of the helix is responsible for the polarization properties of the PHA.

\section{Analytical model of the four-coupled PHA structure}

The unconventional polarization manipulation of the four-PHA structure (Fig. 4) relies on the combination of two elliptically polarized waves of opposite handedness, and the control of their respective intensities. Such a configuration can be simply modeled by the interference of two co-propagating plane waves described by parallel polarization ellipses.

Owing to the field projection rules defined by the two pairs of orthogonal aperture nano-antennas, the electric fields $\textbf{E}_1$ and $\textbf{E}_2$ of these two waves can take the following form:
\begin{eqnarray}
\textbf{E}_1 & = & \sin(\theta-\frac{\pi}{4})\left(1,i b_1,0\right) \exp\left[-i (\omega t-w z)\right],\\
\textbf{E}_2 & = & \cos(\theta-\frac{\pi}{4})\left(1,-i b_2,0\right) \exp\left[-i (\omega t-w z)\right].
\end{eqnarray}
The two light waves show the same wave vector $(0,0,w)$. $(x,y,z)$ are the space coordinates, $\omega$ is the angular frequency, and $t$ refers to time. $b_1$ and $b_2$ are two positive constants smaller than 1, called ellipticity factors. 

Circular polarization arises when:
\begin{equation}
\textbf{E}_1+\textbf{E}_2 = K \left(\pm 1,i,0\right) \exp\left[-i (\omega t-w z)\right],
\end{equation}
where $K$ is a constant, thus imposing:
\begin{equation}
\tan(\theta)  =  \left| \frac{1 \pm b_2}{1 \mp b_1} \right|+\frac{\pi}{4},
\end{equation}
Right and left circular polarizations are then obtained for two specific values of $\theta$ that are dependent on the ellipticity factors $b_1$ and $b_2$ of the two initial waves. 

More generally the electric field resulting from the wave combination can be written
\begin{equation}
(E_x, E_y, 0)= \sin(\theta-\frac{\pi}{4}) (1, i b_1, 0 ) + \cos(\theta-\frac{\pi}{4}) (1, -i b_2, 0).
\end{equation} 
We can consider this total field as an elliptically polarized wave whose major and minor radii take respectively the form :
\begin{eqnarray}
a & = & \frac{\sqrt{2}}{2} \left[|E_x|^2+|E_y|^2 + \sqrt{|E_x|^4+|E_y|^4+2 |E_x|^2 |E_y|^2 \cos\left(2 \Delta \xi \right)}\right]^{\frac{1}{2}},\label{eq:2}\\
b & = & \frac{\sqrt{2}}{2} \left[|E_x|^2+|E_y|^2 - \sqrt{|E_x|^4+|E_y|^4+2 |E_x|^2 |E_y|^2 \cos\left(2 \Delta \xi \right)}\right]^{\frac{1}{2}},\label{eq:3}
\end{eqnarray}
where $\Delta \xi$ refers to the phase difference between $E_x$ and $E_y$ \cite{balanis:book}. From Eqs. \ref{eq:2} and \ref{eq:3}, we anticipated the polarization state of the four-PHA emission, as a function of the projection angle $\theta$ (Supplementary Fig. \ref{fig:model}). We considered the particular case where $b_1=b_2$, i.e., two waves of identical ellipticities and opposite handedness. When these two waves are circularly polarized, the angular spacing  $\Delta \theta$ between right and left polarizations is 90$^{\circ}$. The resulting polarization manipulation is similar to that of a rotating polarizer in front of quarter-wave plate. In that case, $\theta$ refers to the projection angle of an incident linearly polarized wave onto the crystalline axes of the retardation plate. 

Polarization control deviates from this well-known conventional configuration when we consider two elliptically polarized waves ($b_1$ and $b_2$ become smaller than 1). Depending on the ellipticity factor of the two initial waves, the angular spacing $\Delta \theta$ between right and left circular polarizations decreases down to 53$^{\circ}$ when $b_1=b_2=0.5$ and 11.5$^{\circ}$ when $b_1=b_2=0.1$. The four-PHA structure would thus lead to a near-switching effect of circular polarization handedness while rotating linear polarization of an incoming light.  By implementing the polarization states shown in Figs. 4(d) and (f) in our model, we found the polarization manipulations described by the solid lines of Figs. 4(c) and (e), respectively.


\newpage

\begin{figure}[ht!]
\centering
\includegraphics[width=0.9\linewidth]{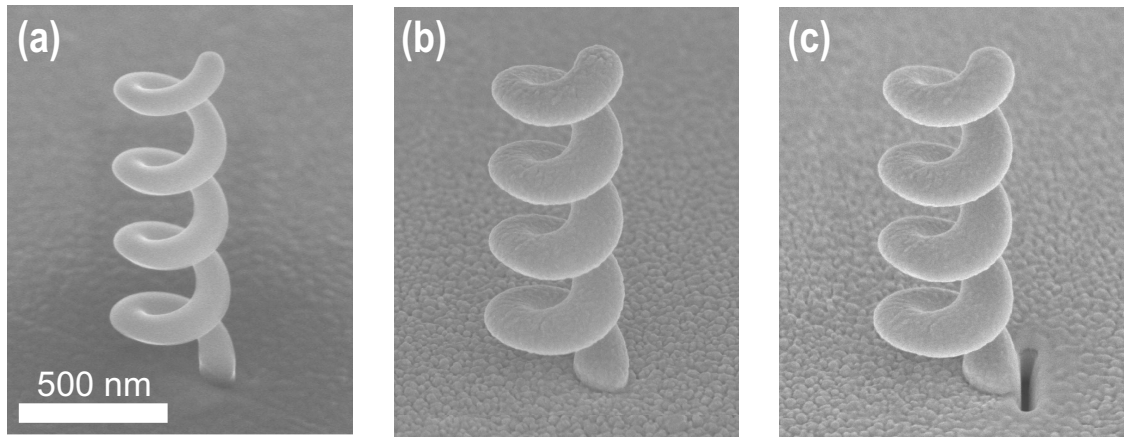}
\caption{\textbf{Fabrication of the plasmonic helical antenna (PHA): three steps.} Scanning electron micrographs of the subwavelength structure after (\textbf{A}) fabrication of the carbon helix skeleton by FIBID \cite{esposito:nl16}, (\textbf{B}) gold deposition onto the helix skeleton, and (\textbf{C}) fabrication of rectangular nano-aperture antenna by FIB milling.}\label{fig:fab}
\end{figure}

\newpage

\begin{figure}[ht!]
\centering
\includegraphics[width=0.6\linewidth]{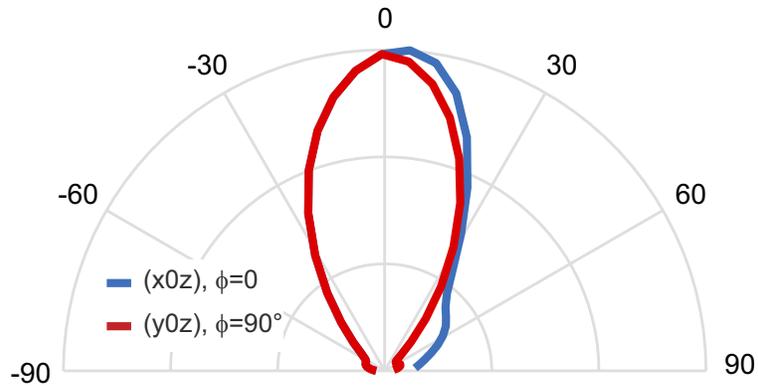}
\caption{\textbf{PHA emission diagram.} The PHA shows a directional emission perpendicularly to the (x0y) ground plane of the antenna, thus confirming its axial operation mode \cite{kraus:book,balanis:book}.}\label{fig:em_diag}
\end{figure}

\newpage

\begin{figure}[ht!]
\centering
\includegraphics[width=1\linewidth]{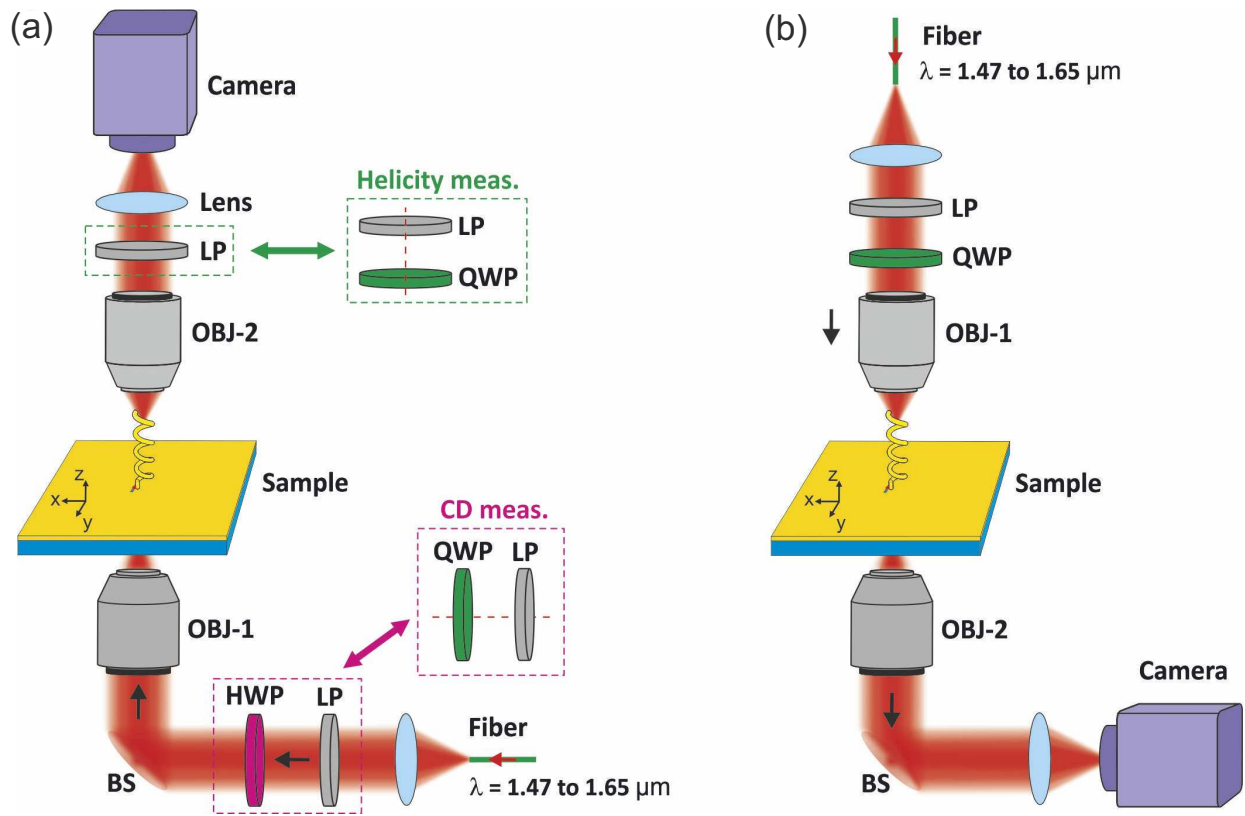}
\caption{\textbf{Experimental setup for optical characterization}. Measurement of (\textbf{A}) the polarization properties of the PHA in emission mode, and (\textbf{B}) the circular dichroism of the structure in collection mode. The beam splitter does not affect the polarization state of the reflected light. BS: Beam Splitter, LP: Linear Polarizer, HWP: Half-Wave Plate, OBJ: Objective, QWP: Quarter-wave plate}\label{fig:bench}
\end{figure}

\newpage

\begin{figure}[ht!]
\centering
\includegraphics[width=0.8\linewidth]{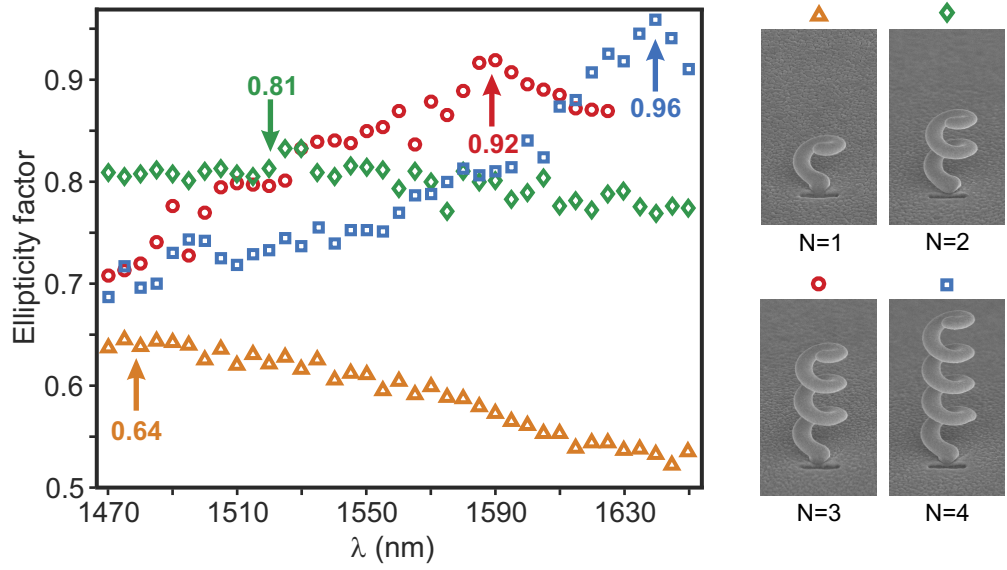}
\caption{\textbf{Optical helicity while varying number of turns in the helix}. Spectrum of the ellipticity factor of the PHA emission for single structures with one turn (orange triangles), two turns (green diamonds), three turns (red circles), and four turns (blue squares)}\label{fig:turns}
\end{figure}

\newpage

\begin{figure}[ht!]
\centering
\includegraphics[width=0.9\linewidth]{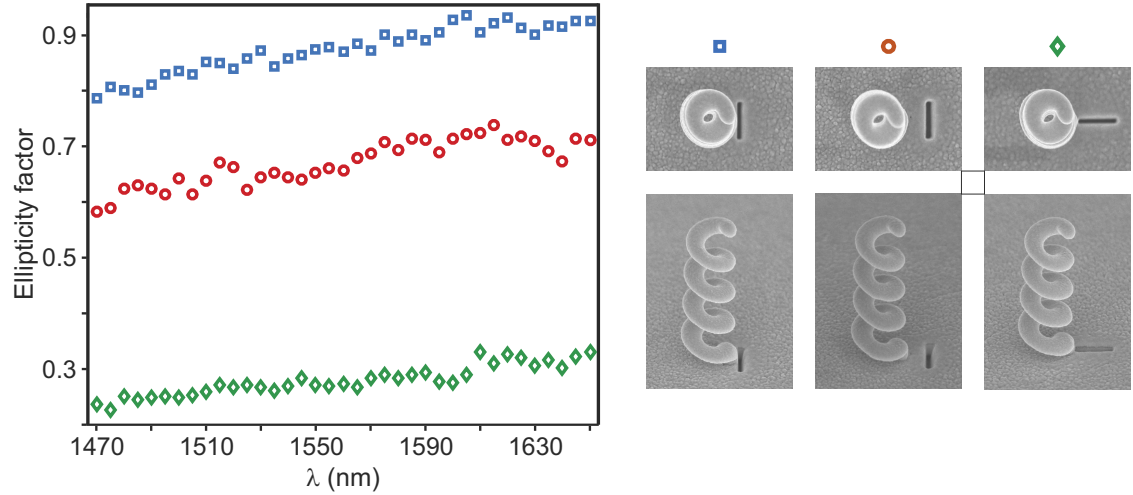}
\caption{\textbf{Circularly polarized emission originates from the excitation of a subdiffraction surface plasmon within the helix}. Spectrum of the polarization ellipticity factor of the PHA emission for a rectangular nano-aperture antenna in contact to the helix pedestal (blue squares), 185 nm away from the helix pedestal (red circles), and turned by 90$^{\circ}$ regarding the two first cases (green diamonds).}\label{fig:end-fire}
\end{figure}

\newpage

\begin{figure}[ht!]
\centering
\includegraphics[width=0.6\linewidth]{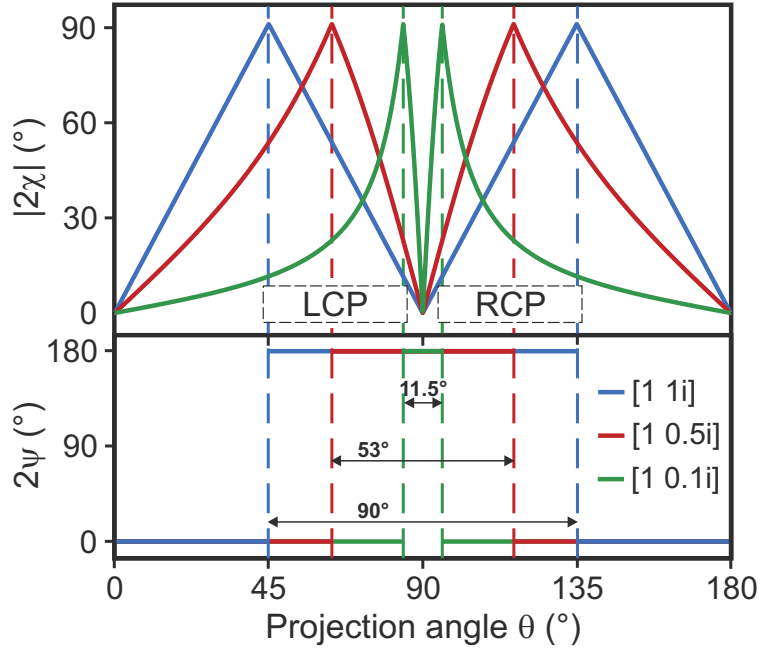}
\caption{\textbf{Theoretical anticipation of the unconventional control of light polarization}. Prediction of the polarization state of the four-PHA emission (i.e., the polarization angle $\left|2 \chi \right|$ and $2 \psi$ on the Poincare sphere), as a function of the projection angle $\theta$ defining the PHA emission intensities. Three ellipticity factors are considered: $b_1=b_2=1$ (blue curves), $b_1=b_2=0.5$ (red curves) and $b_1=b_2=0.1$ (green curves).}\label{fig:model}
\end{figure}

\end{document}